\documentclass [11 pt]{article}

\usepackage{amsmath}
\usepackage{amsfonts}
\usepackage{amssymb}
\usepackage{epsf}
\usepackage{hyperref}

\begin{document}

\begin{center}{\bf{\Large Classical and Quantum Gravitational Collapse in $d$-dim AdS Spacetime}}
{\bf{\Large  I. Classical Solutions}}

\bigskip

{Rakesh Tibrewala$^a$\footnote{e-mail address: rtibs@tifr.res.in}, Sashideep Gutti$^a$\footnote{e-mail address: sashideep@tifr.res.in}, T.P. Singh$^a$\footnote{e-mail address: tpsingh@tifr.res.in} and Cenalo Vaz$^b$\footnote{e-mail address: Cenalo.Vaz@UC.Edu}}

\bigskip

{\it$^a$Tata Insitute of Fundamental Research,}\\
{\it Homi Bhabha Road, Mumbai 400 005, India}

\medskip

{\it$^b$RWC and Department of Physics, University of Cincinnati,}\\
{\it Cincinnati, Ohio 45221-0011, USA}

\end{center}

\begin{abstract}
\noindent
We study the collapse of a spherically symmetric dust distribution in $d$-dimensional AdS spacetime. We investigate the role of dimensionality, and the presence of a negative cosmological constant, in determining the formation of trapped surfaces and the end state of gravitational collapse. We obtain the self-similar solution for the case of zero cosmological constant, and show 
that one cannot construct a self-similar solution when a cosmological constant is included.

\end{abstract}

\section{Introduction}

There are many models of spherical gravitational collapse in classical general relativity       
which exhibit the formation of black holes as well as naked singularities, starting from
regular initial data \cite{si1}, \cite{j1}. The study of quantum effects in the vicinity of the gravitational singularity then becomes significant. Such studies can be divided into two classes : (i) quantum field theory in curved space, and (ii) quantum general relativistic treatment of gravitational
collapse. 

The earliest investigations of quantum field theory in the dynamical background of a collapsing spherical star were probably those due to Ford and Parker \cite{ford} and Hiscock et al. \cite{hiscock}. These works
introduced important techniques, such as the calculation of the quantum flux in the geometric optics
approximation, and the regularization of the 2-d quantum stress tensor, which were used extensively
in later studies. A systematic study of semiclassical effects in gravitational collapse was initiated
by Vaz and Witten in \cite{v1}, \cite{v2} and pursued in a series of papers 
\cite{s1}, \cite{s2}, \cite{s3}, \cite{s4}, \cite{s5}, \cite{si6}, \cite{harada}. Typically, these studies showed an important and interesting difference in the nature of quantum particle creation between the two cases - one in which collapse ends in a black hole, and another in which it ends in a naked singularity. The formation of a black hole is accompanied by the emission of Hawking radiation, as
expected. However, when the collapse ends in a (shell-focusing) naked singularity, there is no evidence of some universal behaviour in the nature of quantum emission. It is typically found though, that the
emitted quantum flux diverges in the approach to the Cauchy horizon. This divergence disappears when the calculation of the quantum flux is terminated about a Planck time before the formation of the Cauchy horizon, when the semiclassical approximation breaks down. Instead of the divergence, one finds that only about a Planck unit of energy is emitted during the semiclassical phase, and a full quantum gravitational treatment of the physics of the singularity and the Cauchy horizon becomes unavoidable. These developments have been reviewed in \cite{si2}.

A full quantum gravitational treatment of collapse can be performed via a midisuperspace quantization
within the framework of quantum general relativity. The aims of such a programme are manifold - to
construct a quantum gravitational description of the black hole; to check if the gravitational
singularity can be avoided in quantum gravity; to obtain a statistical derivation of the black-hole entropy from quantum gravitational microstates; and to determine the role of quantum gravity in ascertaining the nature of quantum emission from a naked singularity. The midisuperspace quantization
programme has been carried out by us in a series of papers 
\cite{wd}, \cite{vawi1}, \cite{vawi2}, \cite{k1}, \cite{k2}, \cite{k3}, \cite{k4},
and work along these lines is still in progress. It is fair to say that while some
progress has been made on aspects related to quantum black holes and black hole entropy, issues related to singularity avoidance and the nature of quantized naked singularities have thus far proved difficult to address, largely bcause of problems relating to finding a suitable regularization scheme for the quantized Hamiltonian constraint in canonical general relativity. Also, we still do not have a definitive answer as to the nature of quantum gravitational corrections to the semiclassical spectrum of Hawking radiation. By this we mean the following: starting from a candidate theory of quantum gravity such as quantum general relativity, one can derive Hawking radiation in the semiclassical approximation using a suitable midisuperspace model. Going beyond the semiclassical approximation, it is expected that quantum gravity will induce (possibly non-thermal) corrections to Hawking spectrum, but this still remains to be worked out in its full generality. It is hoped though that some progress will be possible on these unsolved problems if one makes  
contact with the methods of loop quantum gravity.

All the classical and quantum studies mentioned so far have pertained to gravitational collapse in 3+1 dimensions. Motivated by the desire to overcome some of the obtacles faced in 3+1 physics, we turned attention to investigation of 2+1 gravitational collapse. Homogeneous dust collapse in 2+1 dimensions was first studied in \cite{mann1} and for the case of collapsing shells in \cite{mann2}. This lower dimensional model, though simpler in some aspects, throws up new fascinating issues of its own, which have been studied in the context of inhomogeneous dust collapse in \cite{sashi}, \cite{ss2}, \cite{ss3}.
Classical 2+1 collapse admits a naked singularity for some initial data, but there is no coresponding
quantum particle creation. A black hole solution (the well-known BTZ black hole \cite{btz}) is possible in the presence of a negative cosmological constant, but the thermodynamics and statistics of the quantized BTZ black hole is completely different from that of the 4-d Schwarzschild black hole.

These differences prompt us to the following question: in determining the nature of thermodynamics and
statistics of the quantized black hole, and the nature of quantum emission from naked singularities, what is the role of the cosmological constant, and of the number of spatial dimensions? The present paper is the first in a series
of three papers which addresses this question, by studying classical and
quantum aspects of spherical dust collapse in an AdS spacetime with an arbitrary number of dimensions. In the current paper, we solve the Einstein equations
for a collapsing dust ball in an asymptoticaly AdS spacetime,
and examine the nature of the gravitational singularity. Quantization of
this model will be taken up in two subsequent papers.

The plan of the paper is as follows. In Section II we give results for
spherical gravitational collapse of dust in an asymptotically flat 
$d$-dimensional spacetime. While this problem has been studied earlier by
various authors \cite{ban}, \cite{patil}, \cite{beesham}, \cite{rituparno1}, 
we present here a simpler derivation of the
occurrence of a locally naked singularity, and also obtain new results on the
self-similar solution. More importantly, the results of this Section
serve as a prelude to the corresponding analysis 
presented in Section III, for collapse in an AdS spacetime with arbitrary    
number of dimensions. While gravitational collapse of dust in four dimensional
spacetime with a positive cosmlogical constant has been studied in \cite{ps1},
(see also \cite{ps2}), and for a negative cosmological constant in \cite{mann3}, to the best of our knowledge dust collapse in a $d$-dim
AdS spacetime has not been studied before.

One could question the introduction of a negative cosmological
constant, as is done in this paper, when the observed Universe has  a
cosmological constant which is perhaps positive, or at best zero, but certainly
not negative. Firstly, collapse physics in a deSitter spacetime
is complicated by the presence of a deSitter event horizon, in addition
to the black-hole event horizon. It thus seems natural to first
address the AdS case before moving on to the more realistic, and more 
difficult, deSitter case. There are also
reasons to believe that it would not make sense to directly construct a 
quantum black hole model in a higher dimensional space with a positive 
cosmological constant, because quantum gravity in such a spacetime may not 
exist nonperturbatively \cite{GKS}, \cite{Witten}. Pure 
quantum gravity with a positive cosmological constant may hence not exist 
as an exact theory, but only as a part of a larger system \cite{Witten}.
It is also a question of great interest as to whether studies of statistical
properties of AdS black holes in canonical quantum general relativity can 
benefit from what is known about the AdS/CFT correspondence, as suggested
recently in \cite{vent} for the 4-d case.

\section{Higher Dimensional Spherically Symmetric Dust Collapse in the Absence of a Cosmological Constant}

\subsection {Solution}
The metric for a spherically symmetric space-time can be written in the form
\begin{equation}
ds^{2}=-e^{\mu(t,r)}dt^{2}+e^{\lambda(t,r)}dr^{2}+R^{2}(t,r)d\Omega^{2}
\label{zerometric}
\end{equation}
where 
\begin{equation}
d\Omega^{2}=d\theta_{1}^{2}+\sin^{2}\theta_{1}(d\theta_{2}^{2}+\sin^{2}\theta_{2}(d\theta_{3}^{2}+.......+\sin^{2}\theta_{n-1}d\theta_{n}^{2})).
\end{equation}
Here the number of space-time dimensions is (n+2) where $n\geq1$ is the number of angular coordinates and the 2 designates one time dimension and one radial dimension. For the case where the cosmological constant $\Lambda=0$, Einstein equations are
\begin{equation}
G_{\mu\nu}=kT_{\mu\nu}
\end{equation}
where $k$ is a constant related to Newton's constant of gravitation $G$ (see section (2.4)) and  $T_{\mu\nu}$ is the stress-energy tensor. For the case of non-rotating dust one can choose a synchronous and co-moving coordinate system in which the only non-zero component of the stress-energy tensor is $T_{00}=\epsilon(t,r)$, where $\epsilon(t,r)$ is the energy density of the dust. Further, in co-moving coordinates the $g_{00}$ component of the metric can be chosen to be minus one. With this choice for the metric in (\ref{zerometric}) we get the following independent set of Einstein equations
\begin{eqnarray}
G_{00} &=& \frac{e^{-\lambda}}{R^{2}}\left[-\frac{n(n-1)}{2}R'^{2}+\frac{n}{2}RR'\lambda'+\frac{n(n-1)}{2}e^{\lambda}(1+\dot{R}^{2})+\frac{n}{2}(-2RR''+e^{\lambda}R\dot{R}\dot{\lambda})\right] \nonumber \\
&=& k\epsilon(t,r), \\
G_{01} &=& \frac{n}{2}\frac{(R'\dot{\lambda}-2\dot{R'})}{R}=0, \\
G_{11} &=& \frac{1}{R^{2}}\left[\frac{n(n-1)}{2}\left(R'^{2}-e^{\lambda}(1+\dot{R}^{2})\right)-ne^{\lambda}R\ddot{R}\right]=0, \\
G_{22} &=& -\frac{1}{4}e^{-\lambda}[-2(n-2)(n-1)R'^{2}+2(n-1)RR'\lambda' \nonumber+2(n-2)(n-1)e^{\lambda}(1+\dot{R}^{2}) \nonumber \\
& & -2(n-1)(2RR''-e^{\lambda}(R\dot{R}\dot{\lambda}+2R\ddot{R})+e^{\lambda}R^{2}(\dot{\lambda}^{2}+2\ddot{\lambda})]=0. \\ \nonumber
\end{eqnarray}
Components $G_{33}$, $G_{44}$ etc. are given by expressions similar to that for $G_{22}$ except for overall sine squared factor(s). The Ricci scalar is given by\begin{eqnarray} \label{ricciscalar}
\mathcal{R} &=& \frac{e^{-\lambda}}{2R^{2}}[-2n(n-1)\left(R'^{2}+e^{\lambda}(1+\dot{R}^{2})\right)+2nR\left(R'\lambda'-2R''+e^{\lambda}(\dot{R}\dot{\lambda}+2\ddot{R})\right) \nonumber \\
& & +e^{\lambda}R^{2}(\dot{\lambda}^{2}+2\ddot{\lambda})].
\end{eqnarray}
Solving the equation for $G_{01}$ we obtain 
\begin{equation} \label{elambda}
e^{\lambda}=\frac{R'^{2}}{1+f(r)}.
\end{equation}
In the above expression $f(r)$ is an arbitrary function called the energy function. Integration of the equation for $G_{11}$ after using equation (\ref{elambda}) gives
\begin{equation} \label{Rdot}
\dot{R}^{2}=f(r)+\frac{F(r)}{R^{n-1}}.
\end{equation}
Here $F(r)$ is another arbitrary function and is called the mass function. In what follows we will only consider the so called marginally bound case for which $f(r)=0$. In this case (\ref{Rdot}) can be integrated easily and after choosing the negative sign for the square root corresponding to in-falling matter we get 
\begin{equation} \label{eq for R}
t-t_{c}(r)=-\frac{2}{n+1}\frac{R^{\frac{n+1}{2}}}{\sqrt{F(r)}}
\end{equation} 
where $t_{c}(r)$ is yet another arbitrary function which can be fixed by using the freedom in the choice of the $r$-coordinate. We relabel $r$ such that at $t=0$, $R=r$. With this choice we have
\begin{equation} \label{tc}
t_{c}(r)=\frac{2}{n+1}\frac{r^{\frac{n+1}{2}}}{\sqrt{F(r)}}.
\end{equation}
From the above equations we see that at $t=t_{c}(r)$, $R(t,r)=0$ and this implies singularity formation for the shell labeled $r$ as indicated by the blowing up of the Ricci scalar in (\ref{ricciscalar}).
Finally, substituting for $\lambda$ from (\ref{elambda}) in the equation for $G_{00}$ we find that 
\begin{equation}
k\epsilon(t,r)=\frac{n}{2}\frac{F'}{R^{n}R'}.
\end{equation}
From this one can obtain an expression for the mass function 
\begin{equation} \label{massfunction}
F(r)=\frac{2k}{n}\int\epsilon(0,r)r^{n}dr.
\end{equation}

\subsection{A simple derivation of the naked singularity}

We now look at the nature of the $R=0$ singularity formed at the center $r=0$ of the dust cloud. For this we follow the method used in \cite{barve} and start by assuming that the initial density profile $\epsilon(0,r)$ has the following series expansion near the center $r=0$ of the dust cloud
\begin{equation}
\epsilon(r)=\epsilon_{0}+\epsilon_{1}r+\frac{\epsilon_{2}}{2!}r^{2}+....
\end{equation}
Using this in (\ref{massfunction}) we find that in this case the mass-function 
can be written as
\begin{equation} \label{mass function expansion}
F(r)=F_{n+1}r^{n+1}+F_{n+2}r^{n+2}+F_{n+3}r^{n+3}+......
\end{equation}
where it is to be noted that $n$ is not a free index but, as before, refers to the number of angular dimensions and 
\begin{equation}
F_{n+i}=\frac{2k}{n(n+i)}\frac{\epsilon_{i-1}}{(i-1)!}
\end{equation}
and $i=1,2,3....$ From (\ref{tc}) we know that the singularity curve is given by
\begin{equation} \label{singularity curve}
t_{s}(r)=\frac{2}{n+1}\frac{r^{\frac{n+1}{2}}}{\sqrt{F(r)}}.
\end{equation}
The central singularity at $r=0$ forms at the time
\begin{equation} \label{time for central singularity}
t_{0}=\frac{2}{n+1}\frac{1}{\sqrt{F_{n+1}}}=\frac{2}{n+1}\sqrt{\frac{n(n+1)}{2k\epsilon_{0}}}.
\end{equation}
Here, as a special case, we note that when $\epsilon(r)=\epsilon_{0}$, a constant (Oppenheimer-Snyder collapse), $F(r)=F_{n+1}r^{n+1}$ and the singularity curve is given by 
$t_{s}=2/(n+1)\sqrt{F_{n+1}}$ which is independent of $r$ implying that all shells become singular at the same time as the central shell.
Near $r=0$ one can use the expansion for $F(r)$ as in (\ref{mass function expansion}) and approximate the singularity curve as
\begin{equation} \label{approx singularity curve}
t_{s}(r) \approx t_{0}-\frac{1}{(n+1)}\frac{F_{n+i}}{F_{n+1}^{\frac{3}{2}}}r^{i-1}.
\end{equation}
In the above equation $F_{n+i}$ is the first non-vanishing term beyond $F_{n+1}$ in the expansion for $F(r)$.

One would like to know whether the singularity at $t=t_{0}$, $r=0$ is naked or not, and for this we focus attention on radial null geodesics. We want to check if there are any outgoing radial null geodesics which terminate on the central singularity in the past. Assuming that there exist such geodesics we assume their form  near $r=0$ to be 
\begin{equation} \label{assumed null geodesic}
t=t_{0}+ar^{\alpha}
\end{equation}
Comparing this with (\ref{approx singularity curve}) we conclude that for the null geodesic to lie in the spacetime one must have $\alpha\geq i-1$ and if $\alpha=i-1$ then 
\begin{equation}
a<-\frac{F_{n+i}}{(n+1)F_{n+1}^{\frac{3}{2}}}.
\end{equation}
(This is because $F_{n+i}$ is negative, which will be the case if we demand that $\epsilon(0,r)$ be a decreasing function of $r$). 

Since one is interested in the region close to $r=0$, we expand (\ref{eq for R}) to leading order in $r$ 
to obtain
\begin{equation} \label{approx eq for R}
R \approx r\left[1-\frac{(n+1)}{2}\sqrt{F_{n+1}}\left(1+\frac{1}{2}\frac{F_{n+i}}{F_{n+1}}r^{i-1}\right)t\right]^{\frac{2}{n+1}}.
\end{equation}
From the metric one finds that for null geodesics ${dt}/{dr}|_{NG}=R'$. Differentiating (\ref{approx eq for R}) w.r.t. $r$ we get
\begin{eqnarray} \label{R prime}
R' &=& \left[1-\frac{(n+1)}{2}\sqrt{F_{n+1}}\left(1+\frac{1}{2}\frac{F_{n+i}}{F_{n+1}}r^{i-1}\right)t\right]^{-\frac{(n-1)}{(n+1)}} \nonumber \\
& & \left[1-\frac{(n+1)}{2}\sqrt{F_{n+1}}t-\frac{(n+2i-1)}{4}\frac{F_{n+i}}{\sqrt{F_{n+1}}}r^{i-1}t\right].
\end{eqnarray}
Along the assumed geodesic, $t$ is given by (\ref{assumed null geodesic}). Substituting this in (\ref{R prime}) and equating it with the derivative of (\ref{assumed null geodesic}), i.e. ${dt}/{dr}=\alpha ar^{\alpha-1}$ gives
\begin{eqnarray} \label{key equation}
\alpha ar^{\alpha-1} &=& [1-\frac{(n+1)}{2}\sqrt{F_{n+1}}\left(1+\frac{1}{2}\frac{F_{n+i}}{F_{n+1}}r^{i-1}\right)(t_{0}+a r^{\alpha})]^{-\frac{(n-1)}{(n+1)}} \nonumber \\
& & [1-\frac{(n+1)}{2}\sqrt{F_{n+1}}(t_{0}+a r^{\alpha})-\frac{(n+2i-1)}{4}\frac{F_{n+i}}{\sqrt{F_{n+1}}}r^{i-1}(t_{0}+a r^{\alpha})]. \nonumber \\
\end{eqnarray}
This is the main equation. If it admits a self-consistent solution then the singularity will be naked otherwise not. To simplify this we note that $\sqrt{F_{n+1}}t_{0}=2/(n+1)$, as follows from (\ref{time for central singularity}). 

We first consider the case $\alpha>i-1$. To leading order this gives
\begin{equation} \label{condition on alpha}
\alpha ar^{\alpha-1}=\left(-\frac{F_{n+i}}{2F_{n+1}}\right)^{\frac{2}{n+1}}\left(\frac{n+2i-1}{n+1}\right)r^{\frac{2(i-1)}{n+1}}.
\end{equation}
This equation implies 
\begin{equation}
\alpha=\frac{n+2i-1}{n+1}; \qquad\qquad 
a=\left(-\frac{F_{n+i}}{2F_{n+1}}\right)^{\frac{2}{n+1}}.
\end{equation}
Since $F_{n+i}$ is the first non-vanishing term beyond $F_{n+1}$, we have the condition $i>1$. Also, for consistency we require  $\alpha=(n+2i-1)/(n+1)>i-1$, which together with the previous condition on $i$ implies $1<i<2n/(n-1)$. 

This implies that in 4-dimensions, where $n=2$, we have $1<i<4$, which means that $i=2,3$ are the allowed values. That is, models for which either $\epsilon_{1}<0$ (corresponding to $i=2$) or $\epsilon_{1}=0$, $\epsilon_{2}<0$ (corresponding to $i=3$) will have a naked singularity. 

Similarly in 5-dimensions, where $n=3$, we find that $1<i<3$ implying $i=2$, i.e. only for $\epsilon_{1}<0$ we get naked singularity. In 6-dimensions, $n=4$ and we have 
$1<i<8/3$, implying $i=2$ as the only allowed value, i.e. the singularity is naked only if
$\epsilon_{1}<0$.
One notes that for all higher dimensions $2<2n/(n-1)<3$ and therefore only $i=2$, i.e. $\epsilon_{1}<0$ gives naked singularity.

As another special case we note that for $n=1$, that is in (2+1)dimensions, 
$\alpha=(n+2i-1)/(n+1)=i$ and therefore the condition $\alpha>i-1$ is always satisfied, implying that in this case we always have a naked singularity, which is in agreement with what has been observed in earlier work on (2+1) dimensional dust collapse \cite{sashi}.
 
We next consider the case where $\alpha=i-1$. Here (\ref{key equation}) gives
\begin{eqnarray} \label{eq for critical value} 
(i-1)ar^{i-2} &=& \left(-\frac{(n+1)}{2}\sqrt{F_{n+1}}a-\frac{F_{n+i}}{2F_{n+1}}\right)^{-\frac{(n-1)}{(n+1)}} \nonumber \\
& & \left(-\frac{(n+1)}{2}\sqrt{F_{n+1}}a-\frac{(n+2i-1)}{2(n+1)}\frac{F_{n+i}}{F_{n+1}}\right)r^{\frac{2(i-1)}{(n+1)}}
\end{eqnarray}
which implies $i=2n/(n-1)$. Now the conditions on $i$ are that it be an integer greater than 1. These two conditions are met only for $n=2$ and $n=3$, that is, in (3+1) dimensions and in (4+1) dimensions respectively. For $n=2$, $i=4$ (which corresponds to $\epsilon_{3}<0$) and for $n=3$, $i=3$ (corresponding to $\epsilon_{2}<0$). Since the 4-dimensional case, corresponding to $n=2$, is already reported in the literature \cite{joshi} we focus attention on the 5-dimensional case corresponding to $n=3$.

Substituting $n=3$, $i=3$ in (\ref{eq for critical value}) we obtain
\begin{equation}
8\sqrt{F_{4}}a^{3}+\left(\frac{2F_{6}}{F_{4}}+4F_{4}\right)a^{2}+\frac{4F_{6}}{\sqrt{F_{4}}}a+\frac{F_{6}^{2}}{F_{4}^{2}}=0.
\end{equation}
The above cubic for $a$ has to be solved subject to the constraint 
$0<a<-F_{6}/4F_{4}^{3/2}$ as mentioned earlier. By defining $a=\sqrt{F_{4}}b$ and $F_{6}=F_{4}^{2}\xi$ the above equation is simplified to 
\begin{equation}
2b^{2}(4b+\xi)+(2b+\xi)^{2}=0
\end{equation}
and the constraint on $a$ results in a constraint on $b$ given by $0<b<-\xi/4$. By defining 
$-b/\xi=Y$ and $-1/\xi=\eta$, the above cubic is further simplified to
\begin{equation} \label{cubic equation}
2Y^{2}(4Y-1)+\eta(2Y-1)^{2}=0.
\end{equation}
For a naked singularity to form this equation for $Y$ should have a positive root subject to the constraint $0<Y<1/4$.\\

Now for a general cubic 
\begin{equation}
a_{0}x^{3}+3a_{1}x^{2}+3a_{2}x+a_{3}=0,
\end{equation}
if we define $H\equiv a_{0}a_{2}-a_{1}^{2}$
and $G\equiv a_{0}^{2}a_{3}-3a_{0}a_{1}a_{2}+2a_{1}^{3}$, we have the following conditions on the roots of the cubic \cite{burnside} :
\\

\noindent
1. $G^{2}+4H^{3}<0$, the roots of the cubic are all real. \\
2. $G^{2}+4H^{3}>0$, the cubic has two imaginary roots. \\
3. $G^{2}+4H^{3}=0$, two roots of the cubic are equal. \\
4. $G=0$ and $H=0$, all three roots of the cubic are equal. \\

Using these we can find the conditions on $\eta$ for which the cubic in (\ref{cubic equation}) has at least one real root in the desired range. Here it should be noted that $\eta$ as defined above has to be positive. It is found that for $0<\eta\leq(-11+5\sqrt{5})/4$ all the three roots are real and at least one of these satisfies $0<Y<1/4$. For $\eta>(-11+5\sqrt{5})/4$ the real root is negative. The range of $\eta$ found above implies that for $\xi\leq 4/(11-5\sqrt{5})$ one gets a naked singularity.

We also note that for the Oppenheimer-Snyder collapse mentioned earlier, no naked singularity is formed since all shells become singular at the same time. 

\subsection{Formation of Trapped Surfaces}

We now consider the formation of trapped surfaces. For this consider a congruence of outgoing radial null geodesics with tangent vector $K^{i}=dx^{i}/{dk}$ where $k$ is a parameter along the geodesic and $i=(0,1)$ \cite{tp}. The expansion for these geodesics is given by
\begin{equation} \label{expansion 0 lambda}
\theta=K^{i}_{ ;i}=\frac{1}{\sqrt{-g}}\frac{\partial}{\partial x^{i}}(\sqrt{-g}K^{i}).
\end{equation}
From this one finds that
\begin{equation}
\theta=\frac{nR'}{R}\left(1-\sqrt{\frac{F(r)}{R^{n-1}}}\right)K^{r}.
\end{equation}
Trapping occurs when $\theta=0$ and the above equation with $R'>0$, $R>0$ and $K^{r}>0$ implies that this condition is met for
\begin{equation}
\frac{F(r)}{R^{n-1}}=1.
\end{equation}
In 4-dimensions where $n=2$ we get the well known result
\begin{equation}
\frac{F(r)}{R}=1
\end{equation}
For the general case one finds that the time at which trapping occurs $t_{tr}$ is given by
\begin{equation}
t_{tr}(r)=\frac{2}{n+1}\left(\frac{r^{\frac{n+1}{2}}}{\sqrt{F(r)}}-F(r)^{\frac{1}{n-1}}\right)
\end{equation}
which means that the central shell is trapped at $t_{tr}(r)=2/(n+1)\sqrt{F_{n+1}}$, that is, at the same time as the formation of the central singularity. For the outer shells trapping occurs before those shells become singular.

\subsection{Exterior Solution and Matching with the Interior}

We take the metric in the exterior to be independent of time and given by
\begin{equation} \label{exterior metric}
ds^{2}=-f(x)dT^{2}+g(x)dx^{2}+x^{2}d\Omega^{2}
\end{equation}
where $(T,x,\theta_{1}, \theta_{2}...)$ are the coordinates in the spacetime exterior to the dust cloud. The components of the Einstein Tensor corresponding to the above metric are 
\begin{equation} \label{exterior G00}
G_{00}=\frac{-n(n-1)fg+n(n-1)fg^{2}+nxfg'}{2x^{2}g^{2}},
\end{equation}
\begin{equation} \label{exterior G11}
G{11}=\frac{n(n-1)-n(n-1)fg+nxf'}{2x^{2}f},
\end{equation}
\begin{equation} \label{exterior G22}
G_{22}=\frac{-x^{2}gf'^{2}+2(n-1)(n-2)f^{2}g(1-g)-2(n-1)xf^{2}g'+xf(-xf'g'+2(n-1)gf'+2xgf'')}{4f^{2}g^{2}},
\end{equation}
and $G_{33}$, $G_{44}$ etc. are the related to $G_{22}$ as in the interior.
Solving the vacuum Einstein equations $G_{\mu\nu}=0$ one finds
\begin{eqnarray}
g(x)=\left(1-\frac{C}{x^{n-1}}\right)^{-1}, \\
f(x)=1-\frac{C}{x^{n-1}}
\end{eqnarray}
Here $C$ is a constant of integration. Thus the exterior metric is the Schwarzschild metric
\begin{equation} 
ds^{2}=-\left(1-\frac{C}{x^{n-1}}\right)dT^{2}+\left(1-\frac{C}{x^{n-1}}\right)^{-1}dx^{2}+x^{2}d\Omega^{2}.
\end{equation}
For this to be a valid solution in the exterior we need to match the metric coefficients as well as their first derivatives (extrinsic curvature) in the exterior with the corresponding quantities in the interior at the boundary of the dust cloud $r=r_{s}$ say, \cite{misner}, \cite{toolkit}. This will also determine the only unknown quantity $C$ in the Schwarzchild solution. At the surface the exterior coordinates will be some functions $x=x(t,r_{s})\equiv x_{s}(t)$ and $T=T(t,r_{s})\equiv T_{s}(t)$ of the interior coordinates. These relations imply $dT=\dot{T_{s}}dt$ and $dx=\dot{x_{s}}dt$. Therefore at the surface (where $dr=0$)
\begin{equation}
(ds^{2})_{surf}=\left[-\left(1-\frac{C}{x_{s}^{n-1}}\right)\dot{T_{s}}^{2}+\frac{\dot{x_{s}}^{2}}{1-\frac{C}{x_{s}^{n-1}}}\right]dt^{2}+x_{s}^{2}d\Omega^{2}=-dt^{2}+R_{s}^{2}(t)d\Omega^{2}
\end{equation}
Matching the metric coefficients for $d\Omega^{2}$ gives $x_{s}(t)=R_{s}(t)$ and matching the metric coefficients for $dt^{2}$ then implies
\begin{equation} \label{T dot}
\left(1-\frac{C}{R_{s}^{n-1}}\right)\dot{T_{s}}^{2}-\frac{\dot{R_{s}}^{2}}{1-\frac{C}{R_{s}^{n-1}}}=1.
\end{equation}
To match the extrinsic curvature (second fundamental form) we need the normal to the surface. In the interior coordinates the components of the normal are found to be $n_{\mu}^{i}=(0,R',0....0)$. Similarly in the exterior coordinates the normal is given by $n_{\mu}^{e}=(-\dot{R_{s}},\dot{T_{s}},0...0)$, where the relation $dx-\dot{R_{s}}dt=0$ was used. The extrinsic curvature is given by $K_{ab}=n_{\mu;\nu}e^{\mu}_{a}e^{\nu}_{b}$, where $e^{\mu}_{a}=\partial x^{\mu}/\partial y^{a}$ with $x^{\mu}$ being the coordinates of the (n+2)-dimensional manifold and $y^{a}$ being the coordinates on the boundary of the manifold. Since there is only one undetermined constant $C$, we match only the $K_{\theta_{1}\theta_{1}}$ component of the extrinsic curvature. It can be easily checked that the other components do not give anything new. We find that at the surface the extrinsic curvature in the interior coordinates is given by $K_{\theta_{1}\theta_{1}}^{i}=R_{s}$. Similarly in the exterior coordinates we have $K_{\theta_{1}\theta_{1}}^{e}=R_{s}\left(1-C/R_{s}^{n-1}\right)\dot{T_{s}}$. Equating these two expressions for $K_{\theta_{1}\theta_{1}}$ gives
\begin{equation}
R_{s}=R_{s}\left(1-\frac{C}{R_{s}^{n-1}}\right)\dot{T_{s}}
\end{equation}
Using (\ref{T dot}) and $\dot{R_{s}}^{2}=F(r_{s})/{R_{s}^{n-1}}$ (see (\ref{Rdot})) the above equation gives $C=F(r_{s})$, where from (\ref{massfunction}) it is clear that $F(r_{s})$ is proportional to the total mass of the dust cloud. Thus we find that for the metric coefficients and their first derivatives to be continuous across the boundary the metric in the exterior is given by
\begin{equation} \label{exterior metric}
ds^{2}=-\left(1-\frac{F_{s}}{x^{n-1}}\right)dT^{2}+\left(1-\frac{F_{s}}{x^{n-1}}\right)^{-1}dx^{2}+x^{2}d\Omega^{2}
\end{equation}   

Now $F(r_{s})=(2k/n)\int_{0}^{r_{s}}\epsilon(0,r)r^{n}dr$ and we know that mass of the dust cloud is given by $M=\int_{0}^{r_{s}}\epsilon(0,r)dV$ where $dV$ is the volume element of a spherical shell lying between $r$ and $r+dr$ in $(n+1)$ space dimensions. This volume element is given by 
\begin{equation}
dV=\frac{2\pi^{\frac{n+1}{2}}}{\Gamma(\frac{n+1}{2})}r^{n}dr
\end{equation}
Therefore 
\begin{equation}
M=\frac{2\pi^{\frac{n+1}{2}}}{\Gamma(\frac{n+1}{2})}\int_{0}^{r_{s}}\epsilon(0,r)r^{n}dr.
\end{equation}
This implies 
\begin{equation}
\int_{0}^{r_{s}}\epsilon(0,r)r^{n}dr=\frac{M\Gamma(\frac{n+1}{2})}{2\pi^{\frac{n+1}{2}}}.
\end{equation}
Using this we find that the mass function can be written as 
\begin{equation}
F(r_{s})=C=\frac{2k}{n}\frac{M\Gamma(\frac{n+1}{2})}{2\pi^{\frac{n+1}{2}}}.
\end{equation}

One can also find the constant $C$ in the Schwarzchild solution using the weak field limit. For this we assume that Newton's law for gravity holds for any number of dimensions i.e. $\nabla\cdot g=4\pi G\epsilon(r)$ and $g=-\nabla \phi(r)$. Here $g$ is the gravitational field strength and $\phi$ is the gravitational potential (note: Newton's gravitational constant $G$ being dimensionful will be different in different dimensions, however, this does not affect the form of the equations). Using this we find that in $(n+1)$ spatial dimensions the gravitational potential is given by 
\begin{equation}
\phi(r)=\frac{4\pi GM\Gamma(\frac{n+1}{2})}{2(n-1)\pi^{\frac{n+1}{2}}r^{n-1}}
\end{equation}
where $n>1$ (potential has a logarithmic dependence on $r$ in $2+1$ dimensions). In the weak field limit the Schwarzchild solution is $g_{00}\rightarrow -\left(1-\frac{C}{r^{n-1}}\right)$ and $g_{11}\rightarrow \left(1+\frac{C}{r^{n-1}}\right)$. Also, using geodesic equation we find that generically, in the weak-static field limit $g_{00}=-(1-2\phi)$ and $g_{11}=(1+2\phi)$. Comparing the two expressions for $g_{00}$ (or for $g_{11}$) one finds that $C=2\phi r^{n-1}$ and using the expression for $\phi(r)$ as found above one gets 
\begin{equation}
C=\frac{4\pi GM\Gamma(\frac{n+1}{2})}{(n-1)\pi^{\frac{n+1}{2}}}.
\end{equation}
This expression for $C$ will be the same as that found above from matching if the constant in Einstein's equations is chosen to be $k=4n\pi G/(n-1)$. For $n=2$ this reduces to the value $8\pi G$ as used in 4-dimensional theory and which when used in (\ref{exterior metric}) results in the 
familiar Schwarzchild solution
\begin{equation}
ds^{2}=-\left(1-\frac{2GM}{x}\right)dT^{2}+\left(1-\frac{2GM}{x}\right)^{-1}dx^{2}+x^{2}d\Omega^{2}.
\end{equation}

\subsection{The Self-Similar Solution}

To see the effect of dimensions on the nature of quantum particle flux (which will be described in a work subsequent to this), we would like to have a globally naked singularity. It is known that a locally naked self-similar solution is also globally naked \cite{ori}, where self-similar spacetimes are defined by the existence of a homothetic Killing vector field. Therefore here we look at the dependence on dimensions of the self-similar dust model. 

In a self-similar collapse any dimensionless quantity made from the metric functions,
has to be a function only of $t/r$. This can be seen by starting from the definition 
of a homothetic Killing vector field $\xi$,
\begin{equation}
\xi_{\alpha;\beta}+\xi_{\beta;\alpha}=2g_{\alpha \beta}
\label{homothetic}
\end{equation}
(We emphasize here that we are dealing only with self-similarity of the first kind, which
is defined by the above equation). This condition implies that
\begin{equation}
\mathcal{L}_{\xi}G_{\mu\nu}=0.
\label{Eincollimation}
\end{equation}
where $G_{\mu\nu}$ is the Einstein tensor for the spacetime \cite{HarMae}, \cite{Carr}.
It follows then that the the energy momentum tensor should also satisfy the equation
\begin{equation}
\mathcal{L}_{\xi}T_{\mu\nu}=0.
\label{emtensor2}
\end{equation}
For a perfect fluid the energy-momentum tensor is $T_{\mu\nu}=(p+\mu)u_{\mu}u_{\nu}+pg_{\mu\nu}$. The condition (\ref{emtensor2}) implies the following,
\begin{equation}
\mathcal{L}_{\xi}u^{\mu}=-u^{\mu}
\label{em1}
\end{equation}
\begin{equation}
\mathcal{L}_{\xi}\mu=-2\mu
\label{wm2}
\end{equation}
\begin{equation}
\mathcal{L}_{\xi}p=-2p
\label{em3}
\end{equation}
Based on the above equations, the existence of a homothetic vector implies that the metric components in a comoving coordinate system will be of the form such that the dimensionless quantities become functions of $r/t$, \cite{cahill}. Following \cite{cahill} we assume a general spherically symmetric ansatz of the form
\begin{equation}
ds^2=-e^{2\Phi}dt^2+e^{2\Psi}dr^2+R^2d\Omega ^2
\label{metric}
\end{equation}
and also we have 
\begin{equation}
\mathcal{L}_{\xi}u^{\mu}=-u^{\mu}
\label{ssu1}
\end{equation}
If we assume the vector $\xi$ has only $r$ and $t$ components given by $\xi^{\mu}=\alpha \delta^{\mu}_{r}+\beta \delta^{\mu}_{t}$
and expand (\ref{homothetic}) we get PDEs for the vector components. The condition of comoving metric implies the condition that $\alpha_{,t}=0$ and $\beta_{,r}=0$. After redefining the independent variables to $\bar{r}_{,r}= r/\alpha$ and $\bar{t}_{,t}=t/\beta$ we now redefine the dependent variables by
\begin{equation}
\bar{\Psi}=\Psi+\log(\alpha)-\log(\bar{r})
\label{ppsi}
\end{equation}
\begin{equation}
\bar{\Phi}=\Phi+\log(\beta)-\log(\bar{r}).
\label{pphi}
\end{equation}
Under the above change of variables the equations (\ref{homothetic}) become,
\begin{equation}
\bar{r}R_{\bar{r}}+\bar{t}R_{\bar{t}}=R
\label{euler1}
\end{equation}
\begin{equation}
\bar{r}\bar{\Psi}_{\bar{r}}+\bar{t}\bar{\Psi}_{\bar{t}}=0
\label{euler2}
\end{equation}
\begin{equation}
\bar{r}\bar{\Phi}_{\bar{r}}+\bar{t}\bar{\Phi}_{\bar{t}}=0
\label{euler3}
\end{equation}
So this shows that we can go to a coordinate system in which the metric functions are functions of $\bar{r}/\bar{t}$ . So $\bar{\Psi}$ and $\bar{\Phi}$ are functions of $z$ and $R$ is $r\mathcal{R}(z)$.

It can now be shown that for spherically symmetric, self-similar dust collapse, the mass function is given by $F=\lambda r^{n-1}$, where $\lambda$ is a constant. For this we start with the $G_{01}$ component of Einstein equation 
\begin{equation}
G_{01}=\frac{n}{2}\frac{(R'\dot{\lambda}-2\dot{R'})}{R}=0.
\end{equation}
Defining the self-similarity parameter $z=r/t$ and writing $R\equiv r\tilde{R}$, where $\tilde{R}$ and $\lambda$ being dimensionless are functions only of $z$, the above equation can be written as 
\begin{equation}
-\tilde{R}\frac{d\lambda}{dz}+4\frac{d\tilde{R}}{dz}-z\frac{d\lambda}{dz}\frac{d\tilde{R}}{dz}+2z\frac{d^{2}\tilde{R}}{dz^{2}}=0
\end{equation}
where we have used $R'=\tilde{R}+zd\tilde{R}/dz$ and $\dot{R}=-z^{2}d\tilde{R}/dz$. The above equation is solved easily to obtain
\begin{equation}
e^{\lambda}=c(\tilde{R}+z\frac{d\tilde{R}}{dz})^{2}
\end{equation}
where $c$ is a constant of integration and equals one for the marginally bound case (and will therefore be ignored in what follows).
Similarly the $G_{11}$ component of the Einstein equation gives
\begin{equation}
\frac{n(n-1)}{2}\left(\tilde{R}+z\frac{d\tilde{R}}{dz}\right)^{2}-\frac{n(n-1)}{2}\left(\tilde{R}+z\frac{d\tilde{R}}{dz}\right)^{2}\left(1+z^{4}\left(\frac{d\tilde{R}}{dz}\right)^{2}\right)-nz\tilde{R}\left(\tilde{R}+z\frac{d\tilde{R}}{dz}\right)^{2}\left(2z^{2}\frac{d\tilde{R}}{dz}+z^{3}\frac{d^{2}\tilde{R}}{dz^{2}}\right)=0.
\end{equation}
Solving this we obtain
\begin{equation} \label{key eq for ss}
\tilde{R}^{n-1}\tilde{R}'^{2}=\frac{c}{z^{4}}
\end{equation}
where $c$ is a constant of integration.

Now the mass function is given by $F=\dot{R}^{2}R^{n-1}$ and using $\dot{R}=-z^{2}d\tilde{R}/dz$ this can be written as
\begin{equation}
F=z^{4}r^{n-1}\tilde{R}^{n-1}\tilde{R}'^{2}
\end{equation}
Using (\ref{key eq for ss}) in the above equation we obtain $F=cr^{n-1}$ which is the desired result.

With this result (\ref{eq for R}) becomes 
\begin{equation} \label{eq for self similar R}
R^{\frac{n+1}{2}}=\frac{n+1}{2}\sqrt{\lambda r^{n-1}}(\theta r-t)
\end{equation}
Here $\theta=t_{c}/{r}$ is a constant. This is because in self-similar collapse any dimensionless quantity has to be a function only of $t/r$ whereas $t_{c}/r$, being a function only of $r$ (see (\ref{tc})) has to be a constant. We are interested in finding the behavior of density $\epsilon=nF'/2\kappa R'R^{n}$ in the neighborhood of the centre $r=0$. Using (\ref{eq for self similar R}) and $F'=\lambda(n-1)r^{n-2}$ in the expression for density we find 
\begin{equation}
\epsilon=\frac{4n}{\kappa(n+1)}\left[\left(\frac{n+1}{n-1}\right)\theta^{2}r^{2}-\frac{2n}{n-1}\theta tr+t^{2}\right]^{-1}.
\end{equation}
For $r\approx0$ we neglect the second order term in the above equation and obtain
\begin{equation} \label{density in terms of r}
\epsilon=\frac{4n}{\kappa(n+1)t^{2}}\left[1-\frac{2n}{n-1}\frac{r}{t}\right]^{-1}.
\end{equation}
Also for $r\rightarrow0$, $R^{n+1}=\frac{(n+1)^{2}}{4}\lambda t^{2}r^{n-1}$, which implies 
\begin{equation}
r=\left(\frac{2}{(n+1)t\sqrt{\lambda}}\right)^{\frac{2}{n-1}}R^{\frac{n+1}{n-1}}.
\end{equation}
Substituting this in (\ref{density in terms of r}) we get
\begin{equation}
\epsilon=\frac{4n}{\kappa(n+1)t^{2}}\left[1+\frac{2n\theta}{n-1}a_{n}^{\frac{2}{n-1}}z^{\frac{n+1}{n-1}}+\left(\frac{2n\theta}{n-1}\right)^{2}a_{n}^{\frac{4}{n-1}}z^{\frac{2(n+1)}{(n-1)}}....\right].
\end{equation}
where $a_{n}\equiv2/(n+1)\sqrt{\lambda}$ and $z\equiv R/t$. This shows how the density profile should depend on the number of dimensions to obtain a self-similar solution. The above form for density profile implies that in 4-dimensions ($n=2$) and in 5-dimensions ($n=3$) the self-similar solution corresponds
to an analytic density profile whereas in higher dimensions the density profile is no longer analytic. 

Here we also note that for $n=1$, that is in $2+1$ dimensions, $F=\lambda$ and is thus independent of $r$ and therefore one requires that energy density $\epsilon$ should be zero. Thus self-similarity in $2+1$ dimensions is inconsistent with the presence of matter.

\section{Spherically symmetric inhomogeneous dust collapse in the presence of a negative cosmological constant}

\subsection{Solution}

In the presence of a cosmological constant $\Lambda$, Einstein equations are given by
\begin{equation}
G_{\mu\nu}+\Lambda g_{\mu\nu}=kT_{\mu\nu}.
\label{eilambda}
\end{equation}
For the case $\Lambda<0$ we take $\Lambda\rightarrow-\Lambda$ in which case the Einstein equations become $G_{\mu\nu}-\Lambda g_{\mu\nu}=kT_{\mu\nu}$, where now $\Lambda>0$. The expressions for the components of Einstein tensor are still the same as in the $\Lambda=0$ case. In particular since $g_{01}=0$, therefore we again have
\begin{equation}
G_{01}=\frac{n}{2}\frac{R'\dot{\lambda}-2\dot{R'}}{R}=0.
\end{equation}
The solution of this equation is again given by (\ref{elambda}) and we again consider the marginally bound case so that $f(r)=0$. The 1-1 component of Einstein equations is
\begin{equation}
\frac{1}{R^{2}}\left[\frac{n(n-1)}{2}\left(R'^{2}-e^{\lambda}(1+\dot{R}^{2})\right)-ne^{\lambda}R\ddot{R}\right]-\Lambda R'^{2}=0.
\end{equation}
Integration of this equation gives
\begin{equation}
\dot{R}^{2}=-\frac{2\Lambda}{n(n+1)}R^{2}+\frac{F(r)}{R^{n-1}}
\end{equation}
where as before $F(r)$ is the mass function. Integrating this equation after taking the negative sign for the square root (to account for in-falling matter) we get
\begin{equation} \label{equation for Rlambda}
t-t_{c}(r)=-\frac{2\sin^{-1}\sqrt{\frac{2\Lambda}{n(n+1)}\frac{R^{n+1}}{F}}}{(n+1)\sqrt{\frac{2\Lambda}{n(n+1)}}}.
\end{equation}
Relabeling the $r$ coordinate as in the previous case so that at $t=0$, $R=r$ we get
\begin{equation} \label{tclambda}
t_{c}(r)=\frac{2\sin^{-1}\sqrt{\frac{2\Lambda}{n(n+1)}\frac{r^{n+1}}{F}}}{(n+1)\sqrt{\frac{2\Lambda}{n(n+1)}}}.
\end{equation}
For $t=t_{c}(r)$ we again get $R(t,r)=0$ corresponding to the singularity formation for shell labeled $r$. From $G_{00}-\Lambda g_{00}=k\epsilon(t,r)$ we obtain an expression for $F(r)$ which is again given by (\ref{massfunction}).

\subsection{A simple derivation of the naked singularity}

As before we want to see if null geodesics can come out of the singularity. For this we proceed as before assuming that the density profile near the center is given by
\begin{equation}
\epsilon(r)=\epsilon_{0}+\epsilon_{1}r+\epsilon_{2}\frac{r^{2}}{2!}+...
\end{equation}
From the form of the mass function $F=\frac{2k}{n}\int \epsilon(0,r)r^{n}dr$ we have 
\begin{equation} \label{series mass function lambda}
F(r)=F_{n+1}r^{n+1}+F_{n+2}r^{n+2}+....
\end{equation}
where $F_{n+i}=\frac{2k}{n(n+i)}\frac{\epsilon_{i-1}}{(i-1)!}$.
From (\ref{equation for Rlambda}) and (\ref{tclambda}) we see that the singularity curve is given by
\begin{equation}
t_{s}(r)=\frac{2}{n+1}\frac{\sin^{-1}\sqrt{\frac{2\Lambda}{n(n+1)}\frac{r^{n+1}}{F(r)}}}{\sqrt{\frac{2\Lambda}{n(n+1)}}}.
\end{equation}
This implies that the central singularity at $r=0$ forms at time
\begin{equation}
t_{0}=\sqrt{\frac{2n}{(n+1)\Lambda}}\sin^{-1}\sqrt{\frac{2\Lambda}{n(n+1)F_{n+1}}}. 
\end{equation}
We again note that, as in the $\Lambda=0$ case, when $\epsilon$ is a constant all shells become singular at the same time as the central shell.

We now rewrite the expression for the singularity curve as
\begin{equation}
\sin\left[\sqrt{\frac{(n+1)\Lambda}{2n}}t_{s}(r)\right]=\sqrt{\frac{2\Lambda}{n(n+1)}\frac{r^{n+1}}{F(r)}}.
\end{equation}
It is reasonable to assume that for shells near $r=0$ the time for singularity formation is close to the time for the central shell to become singular i.e. $t_{s}(r)\approx t_{0}$ and we can therefore write $t_{s}(r)=\Delta t_{s}(r)+t_{0}$ where because of the assumption made $\Delta t_{s}(r)\approx 0$. Using this we expand the left hand side of the above equation using the addition formula for sines and make use of $\lim_{x\rightarrow0}\sin(x)=x$ and $\lim_{x\rightarrow0}\cos(x)=1$ to get
\begin{equation}
\sqrt{\frac{(n+1)\Lambda}{2n}}\cos\left[\sqrt{\frac{(n+1)\Lambda}{2n}}t_{0}\right]\Delta t_{s}(r)=-\sin\left[\sqrt{\frac{(n+1)\Lambda}{2n}}t_{0}\right]\frac{F_{n+i}}{2F_{n+1}}r^{i-1}.
\end{equation}
Here $F_{n+i}$ is the first non-zero term beyond $F_{n+1}$ and is negative since we assume a decreasing density profile. Using $\Delta t_{s}(r)=t_{s}(r)-t_{0}$ in the above equation we can finally write the expression for singularity curve for shells near the center as
\begin{equation} \label{singularity curve for lambda}
t_{s}(r)=t_{0}-\sqrt{\frac{2n}{(n+1)\Lambda}}\tan\left[\sqrt{\frac{(n+1)\Lambda}{2n}}t_{0}\right]\frac{F_{n+i}}{2F_{n+1}}r^{i-1}.
\end{equation}
To know whether the central singularity at $t=t_{0}$, $r=0$ is naked or not we focus attention on radial null geodesics and check if there are any outgoing radial null geodesics which terminate on the central singularity in the past. We proceed as in the earlier case, assuming that there exist such geodesics and take their form near $r=0$ to be 
\begin{equation} \label{assumed geodesic}
t=t_{0}+ar^{\alpha}
\end{equation}
where, comparing with (\ref{singularity curve for lambda}), we see that $\alpha\geq i-1$ and if $\alpha=i-1$ then 
\begin{equation}
a<-\sqrt{\frac{2n}{(n+1)\Lambda}}\tan\left[\sqrt{\frac{(n+1)\Lambda}{2n}}t_{0}\right]\frac{F_{n+i}}{2F_{n+1}}
\end{equation}
for the assumed geodesic to lie in the spacetime.
We use (\ref{tclambda}) and (\ref{series mass function lambda}) (retaining only the first two non-zero terms in the latter in the $r\approx0$ approximation) in (\ref{equation for Rlambda}) to get
\begin{eqnarray} \label{R near r0}
R^{\frac{n+1}{2}} &=& \sqrt{\frac{n(n+1)}{2\Lambda}rF_{n+1}}\left(1+\frac{F_{n+i}}{F_{n+1}}r^{i-1}\right) \nonumber \\
& & \sin\left[\sqrt{\frac{(n+1)\Lambda}{2n}}(t_{0}-t)-\tan\left(\sqrt{\frac{(n+1)\Lambda}{2n}}t_{0}\right)\frac{F_{n+i}}{F_{n+1}}r^{i-1}\right].
\end{eqnarray}
Near $r=0$, the time $t$ appearing in the geodesic equation satisfies $t\approx t_{0}$ and therefore the argument of the sine function in (\ref{R near r0}) is close to zero and we use the approximation $\sin x\approx x$ obtaining
\begin{eqnarray}
R&=&r[\frac{(n+1)}{2}\sqrt{F_{n+1}}t_{0}-\sqrt{\frac{n(n+1)}{2\Lambda}}\tan\left(\sqrt{\frac{(n+1)\Lambda}{2n}}t_{0}\right)\frac{F_{n+i}}{2\sqrt{F_{n+1}}}r^{i-1} \nonumber \\
& &-\frac{(n+1)}{2}\sqrt{F_{n+1}}t+\frac{(n+1)}{4}\frac{F_{n+i}}{\sqrt{F_{n+1}}}t_{0}r^{i-1}-\frac{(n+1)}{4}\frac{F_{n+i}}{\sqrt{F_{n+1}}}r^{i-1}t]^{\frac{2}{n+1}}. \nonumber \\
\end{eqnarray}
From the form of the metric we know that the radial null geodesics satisfy $dt/dr|_{NG}=R'$. We take the spatial derivative of the above equation, substitute for $t$ from (\ref{assumed geodesic}) and equate the result to the derivative of (\ref{assumed geodesic}) 
\begin{eqnarray} \label{key equation2}
\alpha ar^{\alpha-1} &=& [-\frac{(n+2i-1)}{(n+1)}\sqrt{\frac{n(n+1)}{2\Lambda}}\tan\left(\sqrt{\frac{(n+1)\Lambda}{2n}}t_{0}\right)\frac{F_{n+i}}{2\sqrt{F_{n+1}}}r^{i-1}-\frac{(n+1)}{2}\sqrt{F_{n+1}}ar^{\alpha} \nonumber \\
& & -\frac{(n+2i-1)}{4}\frac{F_{n+i}}{\sqrt{F_{n+1}}}ar^{\alpha+i-1}][-\sqrt{\frac{n(n+1)}{2\Lambda}}\tan\left(\sqrt{\frac{(n+1)\Lambda}{2n}}t_{0}\right)\frac{F_{n+i}}{2\sqrt{F_{n+1}}}r^{i-1} \nonumber \\
& & -\frac{(n+1)}{2}(\sqrt{F_{n+1}}ar^{\alpha}-\frac{F_{n+i}}{2\sqrt{F_{n+1}}}ar^{\alpha+i-1})]^{-\frac{(n-1)}{n+1}} .
\end{eqnarray}  

Consider first the case $\alpha>i-1$. Keeping terms only to lowest order in $r$ we get
\begin{equation}
\alpha ar^{\alpha-1}=\left(\frac{n+2i-1}{n+1}\right)\left[-\sqrt{\frac{n(n+1)}{2\Lambda}}\tan\left(\sqrt{\frac{(n+1)\Lambda}{2n}}t_{0}\right)\frac{F_{n+i}}{2\sqrt{F_{n+1}}}\right]^{\frac{2}{n+1}}r^{\frac{2(i-1)}{n+1}}
\end{equation}
From this we have 
\begin{equation}
\alpha=\frac{n+2i-1}{n+1}; \qquad\qquad a=\left(-\frac{F_{n+i}}{2\sqrt{F_{n+1}}}\sqrt{\frac{n(n+1)}{n(n+1)F_{n+1}-2\Lambda}}\right)^{\frac{2}{n+1}}
\end{equation}
 where we have substituted for $t_{0}$ in the argument of $\tan$. Since the form of $\alpha$ is exactly the same as in the $\Lambda=0$ case we find that the conditions for naked singularity formation are also the same as mentioned after (\ref{condition on alpha}). That is, in $4$-dimensions $(n=2)$, $1<i<4$ implying $i=2, 3$ are the allowed values so that we get naked singularity for $\epsilon_{1}<0$ or for $\epsilon_{1}=0$, $\epsilon_{2}<0$. Similarly in $5$-dimensions $(n=3)$, $1<i<3$ implying that only $i=2$ is allowed so that we get naked singularity only for $\epsilon_{1}<0$. In all higher dimensions we get naked singularity only if $\epsilon_{1}<0$.
 
Again $n=2$, $i=4$ and $n=3$, $i=3$ are critical cases satisfying $\alpha=i-1$. To analyze these we proceed as in the $\Lambda=0$ case.
For $\alpha=i-1$ (\ref{key equation2}) becomes
\begin{eqnarray}
(i-1)ar^{i-2}&=&[-\frac{(n+2i-1)}{(n+1)}\sqrt{\frac{n(n+1)}{2\Lambda}}\tan\left(\sqrt{\frac{(n+1)\Lambda}{2n}}t_{0}\right)\frac{F_{n+i}}{2\sqrt{F_{n+1}}}r^{i-1} \nonumber \\
& & -\frac{(n+1)}{2}\sqrt{F_{n+1}}ar^{i-1}][-\sqrt{\frac{n(n+1)}{2\Lambda}}\tan\left(\sqrt{\frac{(n+1)\Lambda}{2n}}t_{0}\right)\frac{F_{n+i}}{2\sqrt{F_{n+1}}}r^{i-1} \nonumber \\
& & -\frac{(n+1)}{2}\sqrt{F_{n+1}}ar^{i-1}]^{-\frac{(n-1)}{n+1}},
\end{eqnarray}
which after substituting for $t_{0}$ gives
\begin{eqnarray}
(i-1)ar^{i-2}&=&\left[-\frac{F_{n+i}}{2\sqrt{F_{n+1}}}\sqrt{\frac{n(n+1)}{n(n+1)F_{n+1}-2\Lambda}}-\frac{(n+1)}{2}\sqrt{F_{n+1}}a\right]^{-\frac{(n-1)}{n+1}} \nonumber \\
& &\left[-\frac{(n+2i-1)}{(n+1)}\frac{F_{n+i}}{2\sqrt{F_{n+1}}}\sqrt{\frac{n(n+1)}{n(n+1)F_{n+1}-2\Lambda}}-\frac{(n+1)}{2}\sqrt{F_{n+1}}a\right]r^{\frac{2(i-1)}{n+1}}. \nonumber \\
\end{eqnarray}
Equating the power of $r$ on the two sides gives $i=2n/(n-1)$ (as in the $\Lambda=0$ case). Since $i$ should be an integer greater than one we find that these conditions are satisfied only for $n=2$ ($i=4$) and for $n=3$ ($i=3$). 

Consider $n=3$; in this case the above equation can be written as 
\begin{equation}
8\sqrt{F_{4}}a^{3}+\left(2\sqrt{\frac{6}{6F_{4}-\Lambda}}\frac{F_{6}}{\sqrt{F_{4}}}+4F_{4}\right)a^{2}+4\sqrt{\frac{6}{6F_{4}-\Lambda}}F_{6}a+\left(\frac{6}{6F_{4}-\Lambda}\right)\frac{F_{6}^{2}}{F_{4}}=0
\end{equation}
with the constraint that 
\begin{equation}
0<a<-\sqrt{\frac{3}{12F_{4}-2\Lambda}}\frac{F_{6}}{2F_{4}}.
\end{equation}
 If we define $a=\sqrt{F_{4}}b$ and $F_{6}=F_{4}^{\frac{3}{2}}\sqrt{6F_{4}-\Lambda}\xi$ the equation can be written in the simplified form 
\begin{equation}
2b^{2}(4b+\sqrt{6}\xi)+(2b+\sqrt{6}\xi)^{2}=0
\end{equation}
with the requirement that $0<b<-\sqrt{3/8}\xi$. If we further define $Y=-b/\xi$ and $\eta=-1/\xi$ the equation becomes
\begin{equation}
2Y^{2}(4Y-\sqrt{6})+\eta(2Y-\sqrt{6})^{2}=0.
\end{equation}
For a naked singularity to form this equation for $Y$ should have a solution subject to the constraint $0<Y<\sqrt{3/8}$ and $\eta>0$.\\
Using the conditions, as mentioned earlier, for the roots of a general cubic we can find the conditions on $\eta$ for which the above cubic has at least one real root in the desired range. It is found that for $0<\eta\leq\sqrt{6}(-11+5\sqrt{5})/4$ all the three roots are real and at least one of these satisfies $0<Y<\sqrt{3/8}$. For $\eta>\sqrt{6}(-11+5\sqrt{5})/4$ the real root is negative. The range of $\eta$ found above implies that for $\xi\leq4/\sqrt{6}(11-5\sqrt{5})$ one gets a naked singularity. This shows that the critical case is also similar to the $\Lambda=0$ case except that the allowed range for $\xi$ has shifted.

A similar analysis can be carried out for the case where $n=2$ and $i=4$. By defining $a=F_{3}b$ and $F_{6}=F_{3}^{2}\xi\sqrt{3F_{3}-\Lambda}$ one gets a fourth order equation in $b$. If one subsequently defines $Y\equiv -b/\xi$ and $\eta\equiv -1/\xi$ one gets the equation
\begin{equation}
4Y^{3}(3Y-\sqrt{3})-\eta(Y-\sqrt{3})^{3}=0
\end{equation}
with the consistency conditions $0<Y<1/\sqrt{3}$ and $\eta>0$. It is found that the above conditions are satisfied for $0<\eta<(1590-918\sqrt{3})/(-9+5\sqrt{3})\approx0.066642$ or in terms of conditions on $\xi$ we get $\xi\leq(9-5\sqrt{3})/(1590-918\sqrt{3})\approx-15.0056$.

\subsection{Formation of Trapped Surfaces}

As in the $\Lambda=0$ case, we now look at the formation of trapped surfaces. Considering the expansion of outgoing radial null geodesics as in (\ref{expansion 0 lambda}) we find that
\begin{equation}
\theta=\frac{nR'}{R}\left(1-\sqrt{\frac{F(r)}{R^{n-1}}-\frac{2\Lambda R^{2}}{n(n+1)}}\right)K^{r}.
\end{equation}
From this it is seen that the condition for trapping, $\theta=0$, is met when 
\begin{equation}
\frac{F(r)}{R^{n-1}}-\frac{2\Lambda R^{2}}{n(n+1)}=1
\end{equation}
which for $n=2$ (4-dimensions) reduces to the well known result
\begin{equation}
\frac{F(r)}{R}-\frac{\Lambda R^{2}}{3}=1.
\end{equation}
It is easy to see that, as in the $\Lambda=0$ case, for the central shell, trapping coincides with singularity formation.

\subsection{Exterior Solution with a negative Cosmological constant}

As before we take the metric in the exterior to be 
\begin{equation}
ds^{2}=-f(x)dT^{2}+g(x)dx^{2}+x^{2}d\Omega^{2}.
\end{equation}
The components of Einstein tensor are the same as in (\ref{exterior G00})-(\ref{exterior G22}). Solving the vacuum Einstein equations $G_{\mu\nu}-\Lambda g_{\mu\nu}=0$ we find 
\begin{equation}
g(x)=\frac{n(n+1)x^{n-1}}{2\Lambda x^{n+1}+n(n+1)x^{n-1}+Cn(n+1)},
\end{equation}
\begin{equation}
f(x)=1+\frac{C}{x^{n-1}}+\frac{2\Lambda x^{2}}{n(n+1)}.
\end{equation}
With this the metric in the exterior becomes
\begin{equation} \label{exterior metric for negative lambda}
ds^{2}=-\left(1+\frac{C}{x^{n-1}}+\frac{2\Lambda x^{2}}{n(n+1)}\right)dT^{2}+\left(1+\frac{C}{x^{n-1}}+\frac{2\Lambda x^{2}}{n(n+1)}\right)^{-1}dx^{2}+x^{2}d\Omega^{2}.\end{equation}
Here C is a constant of integration which is fixed by matching the exterior solution to the interior solution at the boundary in exactly the same way as for the $\Lambda=0$ case and the result is $C=-F(r_{s})$ where $r_{s}$ is the boundary of the dust cloud. Thus in 4-dimensions the exterior is 
\begin{equation}
ds^{2}=-\left(1-\frac{2GM}{x}+\frac{\Lambda x^{2}}{3}\right)dT^{2}+\left(1-\frac{2GM}{x}+\frac{\Lambda x^{2}}{3}\right)^{-1}dx^{2}+x^{2}d\Omega^{2}.
\end{equation}

\subsection{The absence of a self-similar solution in the Presence of a $\Lambda$}

It is interesting to note that it is not possible to have a self-similar solution in the presence of a cosmological constant. To see this we begin by noting the condition that dimensionless functions made from the metric are functions only of $t/r$ continues to hold. This follows because the cosmological
constant term in the Einstein equations can be absorbed into the energy momentum tensor in the right
hand side, by taking the $\Lambda$-term as a perfect fluid with equation of state $p=-\rho$. 
Eqn. (\ref{emtensor2}) then continues to hold, with the understanding that the contribution of the
cosmological constant is included in the energy-momentum tensor. The remaining argument, leading to the conclusion that $\tilde{R}$ is a function of $z$ then follows. 

Now if we have a self-similar solution then we can write $R=r\tilde{R}$ with $\tilde{R}$ being dimensionless. If we define $\frac{r}{t}\equiv z$ then the condition of self-similarity implies that $\tilde{R}$ being dimensionless should be a function only of $z$. With this if we now consider the equation $G_{11}-\Lambda g_{11}=\kappa T_{11}$ we get
\begin{equation}
-\frac{n(n-1)}{2}z^{4}\left(\frac{d\tilde{R}}{dz}\right)^{2}-nz^{4}\tilde{R}\frac{d^{2}\tilde{R}}{dz^{2}}-2nz^{3}\tilde{R}\frac{d\tilde{R}}{dz}-\Lambda r^{2}\tilde{R}^{2}=0.
\end{equation}
The explicit presence of $r$ in the above equation implies that $\tilde{R}$ cannot be expressed as a function of $z$ alone and thus we do not have a self-similar solution in the presence of $\Lambda$.
 
The same conclusion also follows from Eqn. (\ref{em3}). With dust matter, the only contribution
to pressure is coming from the cosmological constant, and this pressure is constant. The Lie derivative
on the left hand side is thus zero, whereas on the right hand side the presure is non-zero, leading 
to a contradiction and showing that such a Kiling vector field cannot exist. Physically speaking,
the presence of a cosmological constant introduces a length scale which prevents self-similarity.

\section{Conclusions}
We have studied the collapse of inhomogeneous spherically symmetric dust distribution in arbitrary number of space dimensions both in the absence and in the presence of a cosmological constant. From the analysis presented we see that even though naked singularity is allowed in all dimensions there is more freedom on initial conditions for obtaining  naked singularity in 2+1, 3+1 and 4+1 dimensions, both in the absence as well as the presence of a negative cosmological constant. We have also seen that the formation of trapped surfaces is similar in all dimensions with the central shell getting trapped at the same time when it becomes singular. For outer shells trapping occurs before those shells become singular. We also saw explicitly that in the absence of a cosmological constant, globally naked self-similar models can be constructed all dimensions, whereas in the presence of a cosmological constant such a solution cannot be constructed.

In the second paper in this series, we will study quantum field theory on the curved background
provided by the classical solutions presented here, including the emission of Hawking radiation from
an $n$-dim AdS black hole. In a third paper we will carry out a canonical quantization of this model, and also address the issue of black hole entropy, following the methods of \cite{vent}.

\end{document}